**Vapor-Buoyancy Feedback in an Idealized GCM**

Seth Seidel[1] and Da Yang[2]

[1] Department of Land, Air, and Water Resources, University of California, Davis

[2] Department of the Geophysical Sciences, University of Chicago

Corresponding author:

Seth Seidel (sdseidel@ucdavis.edu)

**Abstract**

Humid air is lighter than dry air at the same temperature and pressure because the molecular weight of water vapor is less than that of dry air. This effect is known as vapor buoyancy (VB). In this work we use experiments in an idealized general circulation model (GCM) to show that VB warms the tropical free troposphere and leads to a significant increase in outgoing longwave radiation (OLR). This radiative effect increases with climate warming, causing a negative climate feedback there. We call this the VB feedback. Although this VB feedback was first corroborated in simplified models, it was heretofore unclear whether the VB feedback is significantly impaired by planetary rotation, clouds, or a countervailing water vapor feedback. However, our GCM simulations show that the VB feedback is robust and maintains an appreciable magnitude when considering these factors.

1. **Introduction**

The molar mass of water vapor (18 g/mol) is less than that of dry air (29 g/mol). According to the ideal gas law, humid air is consequently less dense, or more buoyant, than dry air at a given temperature and pressure. This effect is known as vapor buoyancy (VB). VB is represented in the expression for virtual temperature:

$$T_v = T(1 + vq) \tag{1}$$

$T_v$ is the equivalent temperature a parcel of completely dry air must have in order to have the same density as a given parcel of humid air. $T$ is temperature, $q$ is specific humidity, and $v = \frac{M_d}{M_v} - 1 \approx 0.61$ is the virtual parameter. $M_v$ and $M_d$ are the molar masses of water vapor and dry air, respectively. Although, the origin of VB is well-understood, its effect on climate has drawn little theoretical attention. Furthermore, VB is neglected in several state-of-the-art global climate models, causing systematic errors in their atmospheric temperature, boundary-layer clouds, and clear-sky longwave emission (Yang et al. 2022; Yang and Seidel 2023).

Past studies have contended that VB conspires with the dynamics of Earth's tropics to warm the middle troposphere (2 – 6 km) and cause a negative climate feedback in clear skies (Fig. 4 in Yang and Seidel 2020; Fig. 5 in Seidel and Yang 2020). The warming mechanism, described here, is depicted in Fig. 1. Tropical dynamics constrain the free troposphere to a weak buoyancy gradient (WBG) – that is, a weak gradient in virtual temperature (Charney 1963; Sobel et al. 2002). As a result, comparatively dry parts of the atmosphere must be warmer to maintain equal density to the moist, convective region (Yang and Seidel 2020; Bao and Stevens 2021). However, if we neglect VB, the weak buoyancy gradient becomes a weak *temperature* gradient so that the wet and dry regions of the tropics have the same temperature. Since the temperature profile of the tropics is generally set in the humid regions, this means that the atmosphere is overall warmer due to VB.

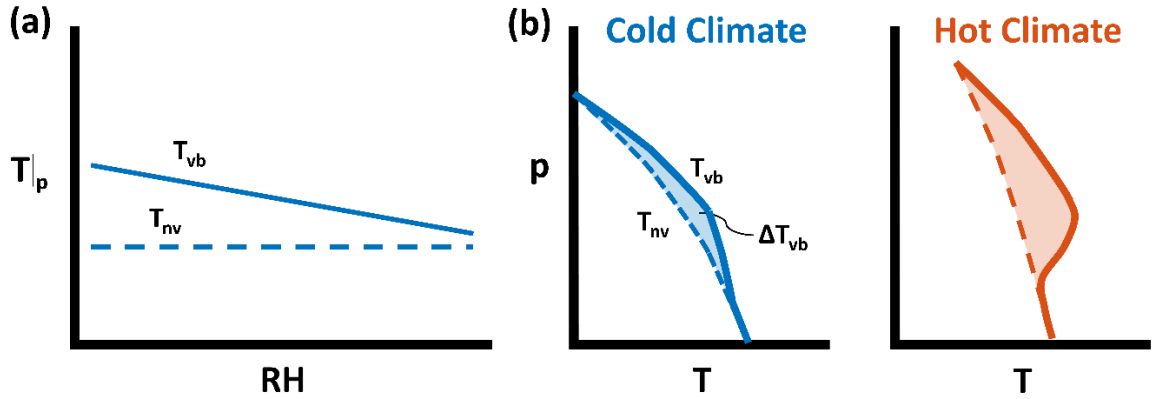

**Figure 1.** (a) The horizontal temperature structure of an atmosphere with VB ($T_{vb}$) and without VB ($T_{nv}$) under WBG dynamics. (b) VB causes the atmosphere to warm more than it otherwise would as $\Delta T_{VB}$ increases with climate warming.

Yang and Seidel (2020) argued that this $\Delta T$ would increase as the climate warms, as the Clausius-Clapeyron scaling of saturation vapor pressure leads to an increase in the water vapor difference between the moist and dry regions of the tropics. This vertically non-uniform warming of the atmosphere causes a lapse-rate-type negative climate feedback, as the ever-warmer middle troposphere emits greater longwave radiation to space. Yang and Seidel (2020) demonstrated this *vapor-buoyancy feedback* (VB feedback) in a simple one-dimensional model of the tropical atmosphere and estimated its magnitude as *O(0.2* W/m²/K*)* in the present-day tropical climate. Seidel and Yang (2020, henceforth "SY20") corroborated the VB feedback mechanism in an idealized two-dimensional cloud-resolving model. Although the SY20 study demonstrated the VB feedback mechanism when convection is explicitly resolved, their idealized experiment had several limitations. First, their model lacked planetary rotation. This left it unclear whether the VB feedback would occur at latitudes for which the WBG assumption is only approximate, as strict

WBG requires negligible planetary rotation. That would confine the VB feedback to a narrow equatorial band, rendering it unimportant in the global energy balance.

Second, the SY20 model relied on the self-organizing nature of convection to give rise to a tropical circulation, whereas a realistic Hadley Circulation may have a different water vapor budget and thereby a distinct moisture distribution. This left it unclear whether the negative VB feedback would be accompanied by a countervailing positive component of water vapor feedback, as might be expected with increasing atmospheric temperature (Held and Shell 2012; Jeevanjee et al. 2021). It was also unclear whether there would be a compensating (positive) or reinforcing (negative) cloud feedback due to vapor buoyancy.

The present study will address these unanswered questions regarding the VB feedback. We shall use simulations in a general circulation model to ask whether rotation or changes in dry-region water vapor significantly affect the VB feedback. We shall use a 1D climate model with explicit line-by-line radiation to test whether the VB feedback can plausibly stabilize hothouse climates.

## 2. Simulation Design

To test whether the VB feedback occurs in an atmosphere with planetary rotation, we rely on aquaplanet simulations in the Community Atmosphere Model, version 6 (CAM6) (Danabasoglu et al. 2020). To calculate radiative fluxes and tendencies, CAM6 employs the Rapid Radiative Transfer Model for GCMs (RRTMG, Mlawer et al. 1997). The model is run with fixed $CO_2$ concentration of 348 ppmv. Shallow convection is parameterized using the Cloud Layers Unified By Binormals (CLUBB) scheme, and deep convection is parameterized using the Zhang-McFarlane scheme (Bogenschutz et al. 2013; Zhang and McFarlane 1995). The prescribed surface temperature is is a meridionally symmetric function of latitude which mimics the observed meridional temperature distribution. The function used is identical to that used in the simulations reported by Yang et al. (2022) The model is run at an approximately 2-degree horizontal resolution.

In order to test the role of VB, we perform three simulations at each surface temperature: (1) CNTL, an unmodified version of the model; (2) MD-DYN, in which VB is removed from the model's dynamics but not its moist parameterizations; and (3) MD-ALL, in which VB is removed from the dynamics as well as the boundary-layer and convection schemes. The MD-DYN experiment is intended to isolate only the physics of the VB feedback, as well as to emulate several climate models who exclude VB from their pressure gradient calculation, but not their moist parameterizations.[1] The MD-ALL simulation is intended to test whether the VB feedback is active compared to a more physically consistent counterfactual. The MD-DYN and MD-ALL experiments correspond to the MD1 and MD2 experiments in Yang et al. (2022); however, we use a different model here.

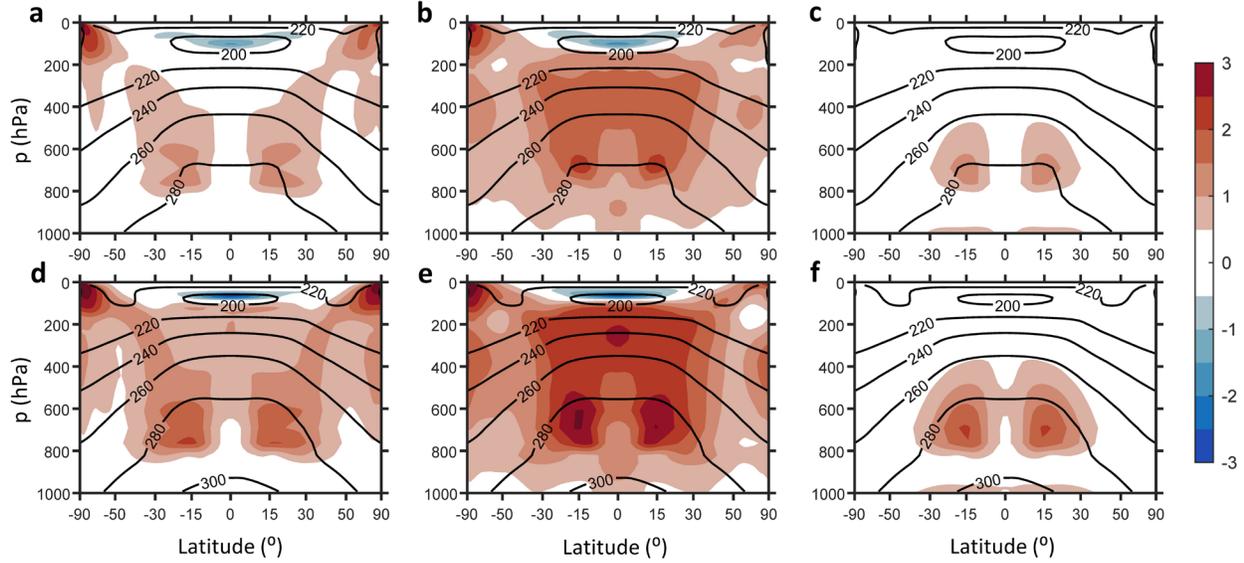

**Figure 2.** Temperature difference due to VB. (a) $\Delta T_{MD-DYN}$ for an equatorial surface temperature of 300 K. (b) $\Delta T_{MD-ALL}$ for an equatorial surface temperature for 300 K. (c) $\Delta T_{theory}$ for an equatorial surface temperature of 300 K. (d) $\Delta T_{MD-DYN}$ for an equatorial surface temperature of 306 K. (e) $\Delta T_{MD-ALL}$ for an equatorial surface temperature of 306 K. (f) $\Delta T_{theory}$ for an equatorial surface temperature of 306 K.

## 3. Results

### 3.1. Hotter atmosphere due to VB

In this section, we want to ask: Is the atmosphere hotter due to VB in an environment with appreciable planetary rotation? Here we are concerned with the difference in temperature $\Delta T$ due to VB:

---

[1] Our design of the MD-DYN experiment is simply to zero out the virtual temperature parameter (*zvir*) which is used in the dynamical core. This affects not only the dynamical core, but also the rest of the model except for the convection and PBL schemes. Notably, this includes surface flux calculations. However, our testing showed these made little difference in the results presented here.

$$\Delta T_{MD-DYN} = T_{CNTL} - T_{MD-DYN}, \qquad (2)$$

$$\Delta T_{MD-ALL} = T_{CNTL} - T_{MD-ALL}, \qquad (3)$$

where $T_{VB}$ is the temperature of the atmosphere in the control simulation, and $T_{MD}$ is the temperature in a mechanism-denial simulation. Figs. 2a and 2b show $\Delta T_{MD-DYN}$ fand $\Delta T_{MD-ALL}$, respectively, at the control surface temperature (300 K at the equator). The subtropical middle troposphere is up to 1 K warmer in the CNTL simulation than the MD-DYN simulation and up to 2.5 K warmer in the CNTL simulation than in the MD-ALL simulation. Figs. 2d and 2e show $\Delta T_{sim}$ when the surface temperature is 4 K greater. $\Delta T_{sim}$ increases with climate warming.

To test whether the warmer atmosphere depicted in Fig. 2 is explained by VB, we develop a simple analytic expression for $\Delta T$. Since density is approximately horizontally homogeneous in the tropical free troposphere (Charney 1963; Sobel et al. 2002), we equate the virtual temperature of a parcel of saturated air ($T_{v,sat}$) and the virtual temperature of air in the simulated or observed atmosphere ($T_v$):

$$T_v = T_{v,sat}. \qquad (4)$$

Substituting the definition of virtual temperature:

$$T(1 + vq) = T_{sat}(1 + vq^*(T_{sat})). \qquad (5)$$

We define $\Delta T = T - T_{sat}$ as the temperature difference due to VB. We further linearize $q^*$ around T using the Clausius Clapeyron relation $\partial_T q^* = \frac{Lq^*}{R_v T^2}$, where L is the latent heat of vaporization of water vapor, and $R_v$ is the gas constant of water vapor. Substituting $T_{sat} = T - \Delta T$ into Eq. 5 and expanding around $\Delta T$:

$$T(1 + vq) = T(1 + vq^*) - \Delta T \left(1 + vq^*(T) + v\frac{L}{R_v T}q^*(T)\right) + \Delta T^2 v \frac{L}{R_v T^2} q^*(T). \qquad (6)$$

We note that $vq^*(T) \ll 1$ and exclude that term. We also exclude the higher order ($\Delta T^2$) term and then reorganize:

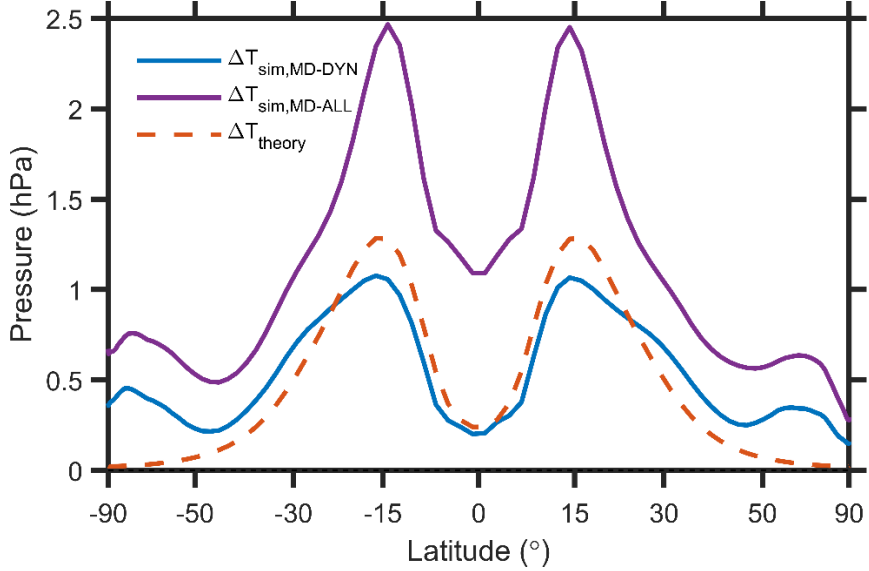

**Figure 3.** Temperature difference due to VB at the 691 hPa model level.

$$\Delta T_{theory} = \frac{vT(q^*(T) - q)}{1 + v\left(\frac{L}{R_v T}\right)q^*(T)}. \qquad (7)$$

Figs. 2c and 2f show $\Delta T_{theory}$ calculated from the control simulation. In the middle troposphere, $\Delta T_{theory}$ is a close match to $\Delta T_{MD-DYN}$ (Figs. 2a and 2d), but substantially underestimates $\Delta T_{MD-ALL}$. This is because $\Delta T_{theory}$ captures the warming due to VB's interaction with the large-scale dynamics, but it does not account for differences due to boundary-layer turbulence or convection. For greater clarity, Fig. 3 compares $\Delta T_{MD-DYN}$, $\Delta T_{MD-ALL}$, and $\Delta T_{theory}$ at the 691 hPa model level. $\Delta T_{theory}$ closely approximates $\Delta T_{MD-DYN}$, particularly in the tropics (± 30º).

Fig. 3 shows that $\Delta T_{theory}$ captures the gross features and scale of the meridional pattern of $\Delta T_{MD-ALL}$ within the tropics. However, in both Fig. 2 and Fig. 3, it is apparent that $\Delta T_{MD-ALL}$ is considerably greater than $\Delta T_{MD-DYN}$. This may be due to differences in temperature and water vapor in the equatorial boundary layer. Figure 4a shows moist static energy is greater in CNTL than in MD-ALL between 800 hPa and the surface. According to the convective quasi-equilibrium hypothesis, upon which the Zhang- McFarlane convection scheme is formulated, we expect that the more energetic CNTL boundary layer would lead to a greater equilibrium temperature in the free troposphere (Zhang and McFarlane 1995; EMANUEL et al. 1994). Then, those free-troposphere temperature differences would be communicated to the rest of the tropics via weak-buoyancy-gradient dynamics. We support the first portion of this hypothesis by calculating a moist adiabatic temperature profile. Using the analytical theory provided by Romps (2017), we first calculate the lifting condensation level for an equator-average (±5°) parcel lifted from 913 hPa model level. Then we calculate the moist adiabatic temperature profile $T_{ma}$ from that lifting condensation level.

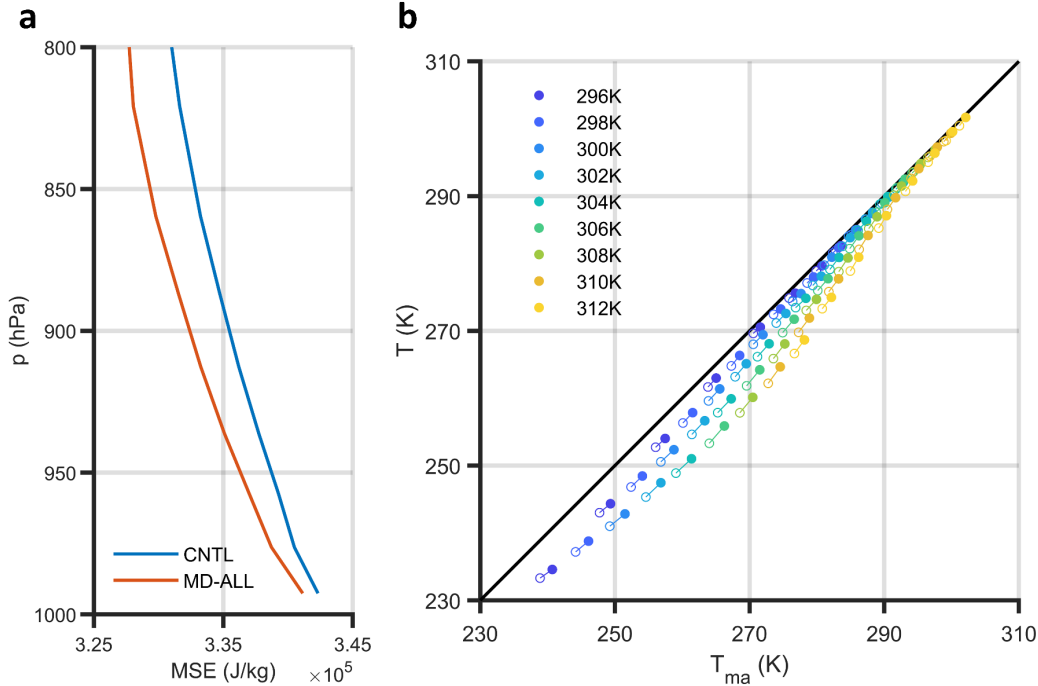

**Figure 4.** (a) Equatorial region (±5°) moist static energy for the CNTL and MD-ALL simulations with an equatorial surface temperature of 300 K. (b) Equatorial region free-troposphere temperature as predicted by moist-adiabatic ascent from the equatorial boundary layer (horizontal axis) plotted against the simulated difference in temperature. Filled circles denote data from the CNTL simulation, and open circles denote MD-ALL.

The horizontal axis of Fig. 4b shows the calculated moist-adiabatic temperature $T_{ma}$ for each model level between 300 and 800 hPa in the same ±5° latitude band. The vertical axis is the actual simulated temperature at that level. Filled circles denote the CNTL simulation, and open circles denote MD-ALL. Overall, $T_{ma}$ is a strong predictor of $T$, as the marks are parallel to the one-to-one line. More importantly, a difference in $T_{ma}$ between CNTL and MD-ALL is generally matched by an approximately equal difference in $T$. This suggests that differences in boundary-layer moist static energy are responsible for the large values of $\Delta T_{MD-ALL}$ at the equator, which are then communicated to the subtropics via large-scale dynamics. However, in the next section we show that the large difference between $\Delta T_{MD-DYN}$ and $\Delta T_{MD-ALL}$ does not lead to large differences in the clear-sky longwave feedback.

Even in a model with appreciable planetary rotation, VB warms the atmosphere by the amount we would expect in theory. This suggests that the vapor-buoyancy feedback is active in the real-world subtropics. We shall evaluate the simulated feedback in the next subsection.

### 3.2. Clear-sky VB feedback

Here we investigate the differences in simulated clear-sky outgoing longwave radiation ($\Delta OLR_{clr}$). The yellow marks in Fig. 5a show tropical-average (± 30° latitude) $\Delta OLR_{clr}$. One may think of this as the total radiative effect of vapor buoyancy. $\Delta OLR_{clr}$ increases as the climate warms, suggesting that the vapor-buoyancy feedback is active there.

We can attribute $\Delta OLR_{clr}$ to simulated differences in tropospheric temperature and specific humidity, as we will show the stratospheric effects are small. To do so, we use clear-sky approximate radiative kernels, which are linear response functions of top-of-atmosphere radiation to perturbations in temperature and humidity. The kernels are described in Appendix B. We take the inner products of the temperature and humidity kernels with the thermodynamic perturbations $\Delta T_{MD-DYN}$ and $\Delta q_{MD-ALL}$, using the lapse-rate tropopause[2] as the upper limit for the integral. The solid blue circles in Fig. 5a show the differences in $\Delta OLR_{clr}$ due to tropospheric $\Delta T_{MD-DYN}$, whereas the red circles show the differences due to tropospheric $\Delta q_{MD-DYN}$. The violet circles show the sum of these two radiative effects, and they closely approximate $\Delta OLR_{clr}$, suggesting that stratospheric effects are negligible. Finally, the open blue circles indicate the temperature radiative effect using $\Delta T_{theory}$.

We can draw two inferences from Fig. 5a. First, VB is responsible for a robust negative feedback as shown by the trend in $\Delta OLR_{clr}$ and that this trend can be explained well by $\Delta T_{theory}$. Second, there is essentially no countervailing change in the water vapor feedback due to VB, as is seen by the fact that tropospheric water vapor does not contribute to the trend in $\Delta OLR_{clr}$. To supplement this result, Appendix A discusses the differences in humidity between CNTL and MD-DYN.

---

[2] We use a lapse-rate threshold of 2 K/km to define the tropopause, as in Zelinka et al. (2020).

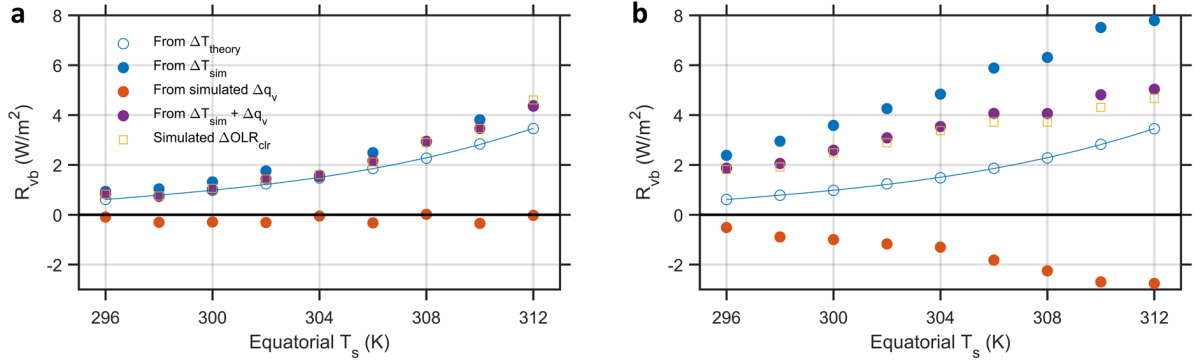

**Figure 5.** Difference in net clear-sky radiation at top of atmosphere for (a) the MD-DYN experiment and (b) the MD-ALL experiment. Closed blue circles indicate the radiative effect implied by $\Delta T_{vb}$ as simulated by either experiment. Open blue circles indicate the radiative effect implied by $\Delta T_{THEORY}$. The red circles indicate the radiative effect implied by differences in specific humidity between the CNTL and MD simulations. The violet circles indicate the sum of the temperature and specific humidity effects. The yellow squares indicate the simulated difference in clear-sky OLR. The blue curve indicates a cubic fit to the open circles.

Figure 5b is the same as Fig. 5a but for the MD-ALL experiment. In this experiment, there *is* an additional positive water vapor feedback due to VB, indicated by the negative trend of the red marks with climate warming. However, this is offset by a greater negative feedback due to $\Delta T_{MD-ALL}$. The sum of the temperature and water vapor radiative effects in the MD-ALL experiment indicate a negative VB feedback similar to that in MD-DYN and to the vapor-buoyancy feedback implied by $\Delta T_{theory}$. The combined temperature and water vapor feedbacks in MD-ALL can be linearly estimated as 0.20 W/m²/K over the whole temperature range, compared to 0.22 W/m²/K for MD-DYN. The MD-DYN and MD-ALL experiments show very similar clear-sky feedbacks. This may be understood by considering that the additional VB-induced warming shown in the MD-ALL experiment (compared to MD-DYN) occurs in convective plumes and communicated to the rest of the tropical atmosphere from there by the large-scale dynamics. The free troposphere's source of water vapor (convection) is warming, so the atmosphere will also be wetter if RH is approximately fixed. This causes a countervailing water vapor feedback which balances the additional temperature feedback in MD-ALL.

The magnitude of a climate feedback is measured by its feedback parameter, which represents a top-of-atmosphere flux sensitivity to a unit increase of surface temperature. We fit a cubic curve to the $\Delta T_{theory}$ marks in Fig. 5 and take its derivative. This gives us the feedback parameter represented by the red curve in Fig. 6. Plotted alongside these data, the blue curve represents the feedback parameter as estimated by a curve fit in the SY20 CRM experiment. The violet curve shows an estimate from a simple 1D column model with two-band radiation, described in Seidel & Yang (2020). The humidity parameter is set to $\beta = 0.5$. These curves represent a hierarchy of models for the VB feedback: a 3D GCM, a 2D CRM, and 1D column model. The three models broadly agree upon the magnitude of the VB feedback and its trend with warming, thus corroborating one another. The differences among these three models are due to differences in virtual temperature profiles, in their humidity profiles, and in their treatments of atmospheric radiation.

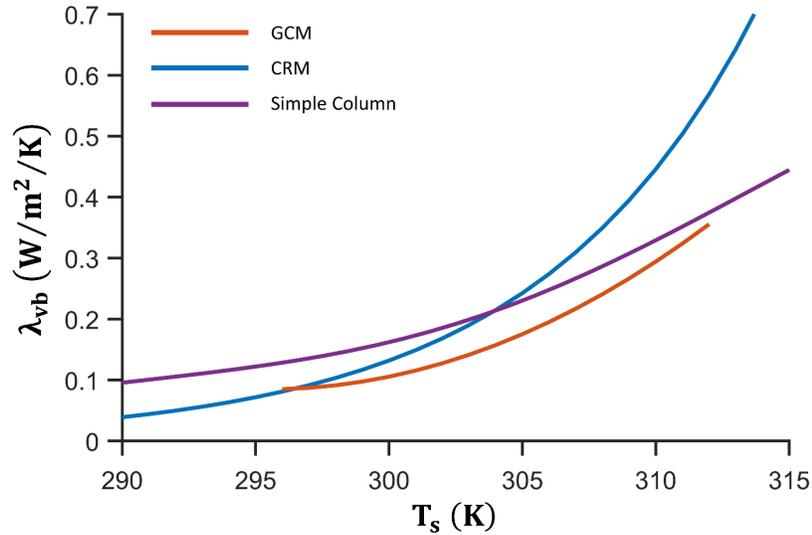

**Figure 6.** Tropical-average VB feedback derived from a hierarchy of climate models. The CRM result is reported in Seidel and Yang (2020), and the Simple Model result is reported in Yang & Seidel (2020).

### 3.3. All-sky and global-average feedbacks due to VB

Our analysis of feedbacks due to VB has so far been limited to clear-sky radiative effects in the tropics. However, it is useful to ask whether VB alters the total climate feedback in a more complete picture: in the global average, with cloud radiative effects included. Figures 7a and 7b show the globally averaged differences in top-of-atmosphere all-sky radiation between CNTL and MD-DYN and CNTL and MD-ALL, respectively. As in the previous sections, we adopt a sign convention that upwelling fluxes are positive so that a positive trend indicates a *negative* climate feedback. For MD-DYN compared to CNTL, the all-sky radiative effect is greater than the clear-sky radiative effect by 4 to 6 W/m². We can roughly estimate the feedback magnitude as 0.35

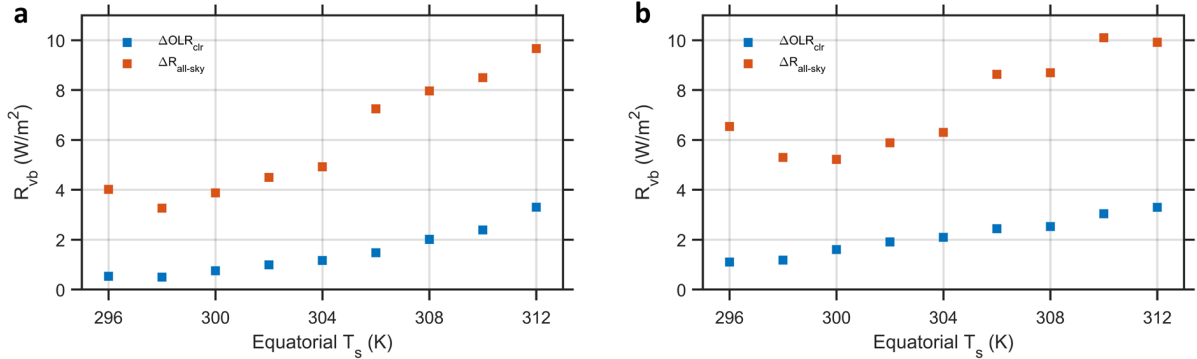

**Figure 7.** Difference in net all-sky radiation at top of atmosphere for (a) the MD-DYN experiment and (b) the MD-ALL experiment (red squares). $\Delta OLR_{clr}$ is shown for comparison.

W/m²/K as the slope from the coldest ($T_{eq} = 296\ K$, $\Delta R_{all} = 4.0\ W/m^2/K$) to the warmest ($T_{eq} = 312\ K$, $\Delta R_{all} = 5.6\ W/m^2/K$) surface temperatures. This is greater than a comparable clear-sky longwave estimate of 0.17 W/m²/K. That is, cloud radiative effects *add* to the vapor buoyancy feedback. Calculating feedbacks in the same way for MD-ALL, the total all-sky feedback is 0.21 W/m²/K, compared to 0.14 W/m²/K for clear-sky OLR. The MD-DYN experiment shows a greater difference in cloud feedback than MD-ALL, which may be attributable to MD-DYN's mismatch between its treatment of large-scale dynamics and its treatment of shallow convection and surface fluxes. However, it should be noted again that the MD-DYN experiment is consistent with several GCMs which appear to exclude VB from their dynamics but not their parameterizations (Yang et al. 2022). Our simulations suggest that that modeling choices may introduce an artificial positive feedback component to the climate system, which may be averted by including VB in the dynamical calculations.

## 4. Discussion

We have used idealized GCM simulations to corroborate a vapor-buoyancy feedback in Earth's tropics. The VB feedback is spatially extensive and robust even when one considers VB's influence on the water vapor feedback. This result was not clear in our earlier study (SY20), which used a cloud resolving model with zero rotation. These results suggest that the VB feedback is physically valid in a warm, wet, rotating atmosphere such as Earth's. Since VB *causes* a difference in clear-sky feedback, the VB feedback is not merely an artifact of unconventional feedback decomposition as suggested by Colman and Soden (2021). Without VB, the mean tropical climate feedback would in fact be less negative. This intuition is important for the development of climate models, as well. Several state-of-the-art GCMs exclude VB from their dynamics and consequently emit less outgoing longwave radiation (Yang and Seidel 2023).

The VB feedback is likely present in those comprehensive climate models which include VB in their pressure gradient calculations. If temperature feedbacks are decomposed under an assumption of constant specific humidity, then the VB feedback would be part of the calculated lapse-rate climate feedback, as it derives from a difference in atmospheric warming relative to surface warming. However, if (temperature-)relative humidity is assumed fixed when calculating feedbacks, as suggested by Held and Shell (2012) and Jeevanjee et al. (2021), the negative VB feedback quantified here would instead be split between a negative lapse-rate feedback and a negative relative humidity feedback (due to decreasing relative humidity). This difficulty could be rectified by redefining relative humidity with respect to a constant-density saturation rather than a constant-temperature saturation, as discussed in Appendix A. However, that assumption may not be appropriate for the boundary layer. There is no perfect choice of moisture variable for feedback decomposition.

The VB feedback represents a considerable improvement in our understanding of hot-climate longwave feedbacks. Recent studies have emphasized the importance of the longwave water vapor window through which longwave emission from the surface may readily escape to space (Koll and Cronin 2018; McKim et al. 2021; Seeley and Jeevanjee 2021; Koll et al. 2023). As the atmosphere becomes hotter and wetter, continuum absorption causes the water vapor window to close, reducing the total clear-sky longwave feedback. Our work suggests that the VB feedback, which strongly increases with surface temperature (Fig. 6), may help to compensate the loss of surface emission and stabilize the climate at hot surface temperatures. In the future it may be fruitful to investigate whether a VB feedback can delay or halt the transition to runaway greenhouse state in hot planetary climates (Ingersoll 1969).

Although this and previous studies have addressed the VB feedback using a wide range of methods and models, future studies may still rectify several remaining knowledge gaps. Further modeling experiments could include continents in order to more realistically simulate VB's influence on cloud feedbacks. It may also be worthwhile to consider the effects of interactive surface temperatures, as the VB feedback's outsize influence in the tropics may reduce the meridional temperature gradient in a warming climate.

*Acknowledgements:*

This work was supported by a National Science Foundation CAREER award a Packard Fellowship for Science and Engineering. Computational resources were provided by the Department of Energy's National Energy Research Scientific Computing Center (NERSC) and by the National Center for Atmospheric Research Computational & Information Systems Lab (NCAR CISL).

**Appendix A: Muted changes in humidity due to VB**

Figs. A1a and A1c show $\Delta RH_T$, the difference in relative humidity between CNTL and MD-DYN. The subscript $T$ denotes that this is a conventional relative humidity, calculated with respect to constant-temperature saturation process. The atmosphere is less humid in the CNTL simulations, particularly in the subtropical middle troposphere, where VB causes the greatest warming. This suggests that VB dries the atmosphere. Furthermore, it appears that the magnitude of $\Delta RH_T$ increases with warming. This is inconvenient for understanding climate feedbacks, as $RH_T$ is often used as the state variable (held fixed) for evaluating temperature-based climate feedbacks (Held and Shell 2012; Zelinka et al. 2020).

For this reason, it is helpful to explore an alternative notion of relative humidity. We define density-relative humidity as the concentration of water vapor in a parcel of air ($q$) compared to the concentration which would be achieved in a constant-density saturation ($q^*(T_v, p)$):

$$\text{RH}_\rho = \frac{q}{q^*(T_v, p)}. \qquad (8)$$

$RH_\rho$ is a useful measure in the free troposphere if one assumes the cloud-free atmosphere is principally moistened by horizontal motions along isentropes (surfaces of constant virtual potential temperature). This alternative definition of relative humidity is not completely novel. Romps (2014) developed an analytical model for tropical relative humidity in which RH was defined as the ratio of specific humidity in a subsaturated environment to the specific humidity of a saturated plume at the same density and pressure, which is how we define $RH_\rho$. That model suggested that RH is enhanced by isentropic mixing from moist plumes but reduced by subsidence drying. However, it neglected the lightness of water vapor, so the distinction between $RH_\rho$ and $RH_T$ was not apparent.

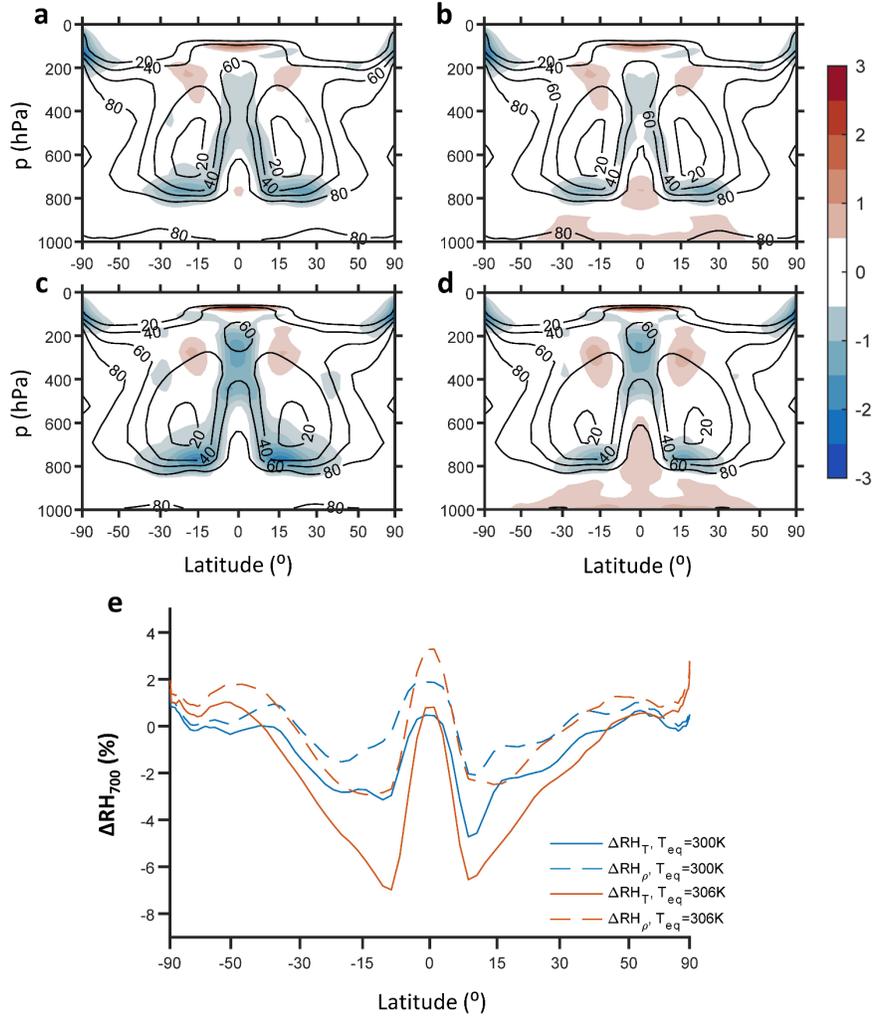

**Figure A1.** Difference in zonal-average relative humidity due to VB (color contours); zonal-average CNTL simulation humidity (black contours). (a) $\Delta RH_T$ for an equatorial surface temperature of 300K. (b) $\Delta RH_\rho$ for an equatorial surface temperature of 300K. (d) $\Delta RH_\rho$ for an equatorial surface temperature of 306K. (e) $\Delta RH_T$ and $\Delta RH_\rho$ at 700 hPa.

In Earth's atmosphere, $RH_\rho$ is necessarily greater than $RH_T$ since adding more water vapor to a parcel under a constant-density process necessitates a lower temperature, and smaller $q^*$. The relationship between $RH_\rho$ and $RH_T$ can be derived from the relationship $q^*(T_v, p) = q^*(T - \Delta T_{theory})$. Linearizing $q^*$ around the Clausius-Clapeyron relation $\partial_T q^* = \frac{L}{R_v T^2} q^*$, this gives:

$$RH_\rho = \frac{q}{q^*(T,p) - \frac{L}{R_v T^2} q^*(T,p) \Delta T_{theory}}. \qquad (9)$$

This simplifies to:

$$RH_\rho = \frac{RH_T}{1 - \frac{L}{R_v T^2} \Delta T_{theory}}, \qquad (10)$$

where $RH_T$ is the conventional (temperature-)relative humidity. The departure between $RH_\rho$ and $RH_T$ grows considerably with warming, owing to its dependence on $\Delta T_{theory}$. For a parcel of air with $RH_T = 50\%$ and a pressure of 800 hPa, $RH_\rho$ increases from 52% at a temperature of 280 K to 57% at a temperature of 300 K.

Figs. A1b and A1d show $\Delta RH_\rho$ for the MD-DYN experiment at two different surface temperature.[3] Since $\Delta T_{theory}$ is zero in an atmosphere without VB, $RH_\rho = RH_T$ for the mechanism-denial experiments. Comparing to Figs. A1a and A1c, $\Delta RH_\rho$ shows weaker minimums than $\Delta RH_T$ in the subtropical middle troposphere. In Fig. A1e, we compare $\Delta RH_T$ and $\Delta RH_\rho$ at the 691 hPa model level, which tells a similar story: $\Delta RH_\rho$ shows weaker VB-induced drying of the subtropics compared to $\Delta RH_T$, and a compensating wetting of the deep tropics. This suggests an intensification or narrowing of the model's intertropical convergence zone due to VB. Furthermore, $RH_\rho$ is more nearly fixed than $RH_T$ is as the climate warms, especially in the subtropics.

**Appendix B. Clear-sky radiative kernels**

To decompose perturbations in clear-sky OLR into contributions due to temperature and water vapor, we construct clear-sky radiative kernels, shown in Fig. A2. Radiative kernels are linear response functions of top-of-atmosphere radiation to atmospheric properties. We use the approximate kernel technique from Cronin and Wing (2017), calculating the linear response of top-of-atmosphere radiation to small perturbations in the zonal mean thermodynamic profiles. Our kernels are derived from +0.5 K perturbations in temperature and -1% perturbations in specific humidity from the CNTL simulation zonal average. We calculate a separate radiative kernel for each surface temperature.

---

[3] For the simulation without VB, $RH_\rho = RH_T$.

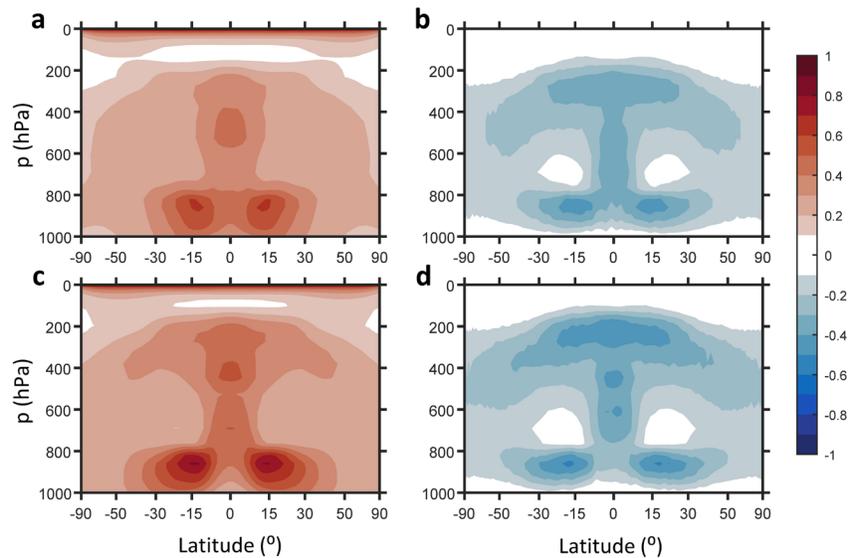

**Figure A2.** Radiative kernels (W/m$^2$/K/100 hPa) for (a) temperature and (b) humidity at an equatorial surface temperature of 300K, and for (c) temperature and (d) humidity for an equatorial surface temperature of 306K. The humidity kernels reflect 1K of warming at fixed $RH_T$.